\documentclass{article}
\usepackage{spconf,amsmath,graphicx}
\usepackage{amsfonts}  
\usepackage{caption}
\usepackage{subcaption}  
\usepackage{balance}  
\usepackage{url}
\usepackage{hyperref}

\newcommand{\maxsnr}{\mathrm{SNR}_\mathrm{max}}

\title{IMPROVING BIRD CLASSIFICATION WITH UNSUPERVISED SOUND SEPARATION
}
\name{Tom Denton, Scott Wisdom, John R.\ Hershey}
\address{Google Research}

\begin{document}
\ninept
\maketitle
\begin{abstract}

This paper addresses the problem of species classification in bird song recordings.  The massive amount of available field recordings of birds presents an opportunity to use machine learning to automatically track bird populations.  However, it also poses a problem: such field recordings typically contain significant environmental noise and overlapping vocalizations that interfere with classification.  The widely available training datasets for species identification also typically leave background species unlabeled.  This leads classifiers to ignore vocalizations with a low signal-to-noise ratio.  However, recent advances in unsupervised sound separation, such as \emph{mixture invariant training}  (MixIT), enable high quality separation of bird songs to be learned from such noisy recordings. In this paper, we demonstrate improved separation quality when training a MixIT model specifically for birdsong data, outperforming a general audio separation model by over 5 dB in SI-SNR improvement of reconstructed mixtures. 
We also demonstrate precision improvements with a downstream multi-species bird classifier across three independent datasets. The best classifier performance is achieved by taking the maximum model activations over the separated channels and original audio. Finally, we document additional classifier improvements, including taxonomic classification, augmentation by random low-pass filters, and additional channel normalization.
\end{abstract}
\begin{keywords}
Source separation, unsupervised learning, classification algorithms, ecology
\end{keywords}
\section{Introduction}
\label{sec:intro}

Machine learning for bioacoustics is a growing area, with potential to transform our understanding of the natural environment. Recent successes include the release of apps like BirdNET and Merlin Sound ID that allow users to identify birds in their immediate surroundings. Meanwhile, groups like Rainforest Connection use audio data to identify critical habitat for endangered species and prevent illegal deforestation and poaching.

Acoustic data provides more opportunities to observe animals than camera traps, which suffer from a narrow field of view and limited range. Birds, whales, and countless other animals use vocalizations as a safe and effective way to communicate and mark territory. These vocalizing animals also act as \emph{indicator species}, allowing further inference about habitat and ecosystem health, including plant and insect food sources, predators, and overall biodiversity. For decades, efforts like the North American Breeding Bird Survey \cite{hudson2017role} have used expert listeners to track bird populations and inform decisions on habitat protection and restoration. The availability of cheap audio recording tools allows vast improvements in sampling coverage but produces quantities of data that require the introduction of automated processing. 

However, bird species classifiers still have difficulty with raw soundscape data, especially when there is significant background noise or during the \emph{dawn chorus}, when vocal activity is at its peak and many species vocalize simultaneously. This difficulty is exacerbated by the fact that available training data for bird song usually carries a label for a single principal species in an audio clip, even when multiple species are vocalizing, and leaves quieter background calls unlabeled. This produces a domain shift when applying classifiers to soundscapes, where we want to identify all species in the recording, not just the most prominent. For example, a very common species like the red-winged blackbird (\emph{rewbla}\footnote{For brevity, we refer to individual species by their 6-letter eBird code.}) may appear very frequently in the background of recordings of other species, leading a classifier to learn that ``quiet'' rewbla vocalizations are usually unlabeled.

Recently there have been significant advances in unsupervised audio source separation using noisy training data. Mixture invariant training (MixIT) \cite{wisdom2020mixit} creates a separation model that can tease apart the individual sounds in a single-channel recording.
Crucially, training a MixIT model does not require clean audio sources, unlike most previous systems. This provides a clear opportunity for bioacoustics: with MixIT separation models, we can isolate individual vocalizations in chaotic soundscapes and suppress excessive background noise, allowing better classification performance.

\textbf{Contributions: }
In this paper, we demonstrate classification improvement when combining an unsupervised separation model with a birdsong classifier. We show improvement on a range of classification metrics across three different evaluation datasets. 

We also demonstrate a number of improvements on the base classification model, relative to previously published efforts. Most notably, we introduce \emph{taxonomic training}, where we train classification heads for each level of a species taxonomy: species, genus, family, and order, along with a binary bird detection head. This allows models to learn higher-level labels before sorting out the sometimes-subtle differences between closely related species. In the context of hierarchical classification \cite{silla2011survey}, we train a \emph{global} classifier, though we are only using the higher taxonomic classes to improve learning of the species problem. We also document gains from additional mel-spectrogram clean-up, low-pass augmentation, and ensembling multiple classifiers.

\section{Methods}
\label{sec:methods}

\subsection{Training Data Activity Detection}

All training data is \emph{weakly labeled}, providing a global label for a principal species in the recording, though there may be long periods where the labeled species does not occur. We preprocess the training datasets with an \emph{activity detector} to select short windows that are likely to contain the labeled species. To choose a window, we create a log mel-spectrogram for each recording and compute the energy of each frame. We then apply a wavelet peak detector \cite{du2006improved} (scipy's \texttt{find\_peaks\_cwt}) to find frames with high energy. We extract a 6-second window around each peak frame and select up to five windows in decreasing order of peak-frame energy. This process reliably finds the labeled species, but occasionally picks a background vocalization or human speech.

\subsection{Separation Model Overview}
\label{ssec:sep_overview}

Recently, mixture invariant training (MixIT) \cite{wisdom2020mixit} was proposed as a means of training sound separation models on in-the-wild acoustic mixtures that lack ground-truth reference sources. The basic idea is to create \emph{mixtures of mixtures} (MoMs) from two training reference mixtures, estimate $M$ sources from the MoM, assign each of these sources to one of the two reference mixtures, and compute a reconstruction loss between each reference mixture and the sum of its assigned sources. The assignment of sources to mixtures is chosen to minimize this reconstruction loss.  The model is then optimized by minimizing the reconstruction loss given the chosen assignment of sources.  The model can, in effect, produce fewer than $M$ output sources by generating near-zero signals for some of the outputs.  

 In more detail, two reference mixtures $x_1\in\mathbb{R}^T$ and $x_2\in\mathbb{R}^T$ are added together to create a MoM,
$
  \bar{x} = x_1+x_2.
$ 
The separation model $f_\theta$ predicts $M$ sources $\mathbf{\hat{s}}\in\mathbb{R}^{M\times T}$ from the MoM:
$
  \hat{\mathbf{s}} = f_\theta(\bar{x})
$. 
Using a discriminative signal-level loss $\mathcal{L}$ between reference mixtures $x_n$ and separated sources $\hat{\bf s}$, the MixIT loss \cite{wisdom2020mixit} estimates a mixing matrix $\mathbf{A}\in\mathbb{B}^{2 \times M}$:
\begin{align}
\mathcal{L}_\mathrm{MixIT}
\left(
    \{{x}_n\}, \hat{\bf s}
\right)
=
\min_{{\bf A}\in\mathbb{B}^{2 \times M}} \, & \sum_{n=1}^2
    \mathcal{L}
    \left(
        {x}_n, [{\bf A} \hat{\bf s}]_n
    \right)
\label{eq:mixit}
\end{align}
where $\mathbb{B}^{2 \times M}$ is the set of $2\times M$ binary matrices where each column sums to $1$. This matrix assigns each separated source $\hat{s}_m$ to one of the reference mixtures ${x}_n$. For the signal-level loss $\mathcal{L}$, we use a negative thresholded SNR loss \cite{wisdom2020mixit, wang2021sequential}:
\begin{equation}
    \mathcal{L}({y}, \hat{y})
    =-10\log_{10}
    \frac{\|{y}\|^2}
    {\|{y}-\hat{y}\|^2 + \tau \|{y}\|^2}
    \label{eq:snr}
\end{equation}
where $\tau=10^{-\maxsnr / 10}$ acts as a soft threshold that clamps the loss at $\maxsnr$.
This threshold prevents examples that are already well-separated from dominating the gradients within a training batch.
We find $\maxsnr=30$ dB to be a good value.



The separation model is a masking-based architecture.
First, the input audio is transformed with a learnable basis \cite{luo2019conv}. These coefficients are fed to
an improved \emph{time-domain convolutional network} (TDCN++) \cite{kavalerov2019universal}, which predicts $M$ masks through a sigmoid activation. These masks are multiplied with the analysis coefficients, and audio waveforms are synthesized using overlap-and-add with a learnable synthesis basis. Finally, a mixture consistency projection \cite{wisdom2018consistency} is applied that constrains sources to add up to the original input.

\subsection{Data Augmentation}
Data augmentation is well known to be a fundamental component of a good birdsong classifier \cite{birdclef2019}. We apply the following augmentations to the training data during classifier training:

    \emph{Random time shift}: We choose a random 5-second training window from each 6-second example in the training set.
    
    \emph{Random gain}: For each training example, we peak normalize to a uniform random value between $0.05$ and $0.75$.
    
    \emph{Example mixing}: With 50\% probability, we mix in a second labeled training example with random gain. The target label set is the union of the two examples' labels.
    
    \emph{Noise mixing}: We use noise recordings from the following sources:
            (1)
            the negative class from the 2018 DCASE Bird Audio Detection Challenge \cite{Berger_JKU},
            (2)
            selected categories from the BBC Nature Sound Effects Library, and
            (3)
            a random set of recordings from the Common Voice dataset \cite{commonvoice}.
        Noise is mixed into 75\% of training examples, with SNR chosen uniformly at random between 0 and 40 dB. With 10\% probability, we suppress labeled audio and labels to train on noise-only.
    
    \emph{Random low pass}: In addition to reducing gain, increasing distance from an audio source has a low-pass effect. We simulate application of a random low-pass filter by scaling the frequency bands in the mel-spectrogram. 
   
   \subsection{Classifier Architecture}

    For the classifier frontend, we compute the mel-spectrogram of the augmented audio and apply PCEN \cite{pcen}, which has been widely observed to help classification \cite{pcen_howwhy}.
    We also apply a final channel-wise normalization to the PCEN mel-spectrogram. For this final normalization, we compute the channel mean and variance and remove all outliers more than one positive deviation from the mean. We then recompute the mean and variance, with outliers excluded, and normalize the channel.
    
    The classifier model uses an EfficientNet-B0 backbone \cite{tan2019efficientnet} applied directly on the PCEN mel-spectrograms. Following the EfficientNet, we apply AutoPool \cite{autopool} to reduce the time dimension and project to a 1280-dimensional hidden space.
    Separate classification heads are used for each level of the species taxonomy (species, genus, family, and order), with labels automatically derived from the species. Losses for the genus, family, and order outputs are weighted at $0.1$. These outputs are only used during training.

    All output heads are trained with binary cross-entropy loss (allowing each label to activate independently), with a label-smoothing factor of $0.1$. 
    Each model is trained five times from scratch, and are used as an ensemble by averaging the output probabilities in the logit domain for each species.

\vspace{-2.5pt}
\subsection{Classifying Separated Audio}
\vspace{-2.5pt}

To combine the separation and classification models, we apply a separation model to an input audio window to obtain $M$ output channels. We then apply the classification model to each separated channel and the original audio and take the maximum probability for each species. We find that including the original mixture as a channel significantly improves the results, as shown in Table~\ref{table:sepclasseval}.

\section{Experiments}
\label{sec:experiments}

\subsection{Birdsong Training Datasets and Processing}
\label{subsec:traindata}

For training both the separation model and classifiers, we use a combination of data from Xeno-Canto \cite{xenocanto} and the Macaulay Library \cite{macaulay}. These source files range from a few seconds to over five minutes in length and are weakly labeled, with a single label for the entire file but no segmentation information. Some Xeno-Canto files also include \emph{background labels} indicating species other than the primary labeled species appearing in the recording. In this case, there is still no segmentation information available, and birds unknown or difficult to identify for the recordist may still be unlabeled.

Each training dataset is specialized to a particular set of target species. 

We select up to 250 training recordings per species, preferring Macaulay recordings and Xeno-Canto recordings with a high user rating and background labels. We additionally include 250 files randomly selected from species outside the target species set. During classifier training, segments from these additional files are provided without a species label, though any relevant taxonomic labeling (genus, family, order) is still included for computing taxonomic loss.

\begin{figure*}[t]
  \begin{subfigure}[b]{0.3\textwidth}
      \centering
      \includegraphics[width=\textwidth,trim={1.49cm 0 0 0},clip]{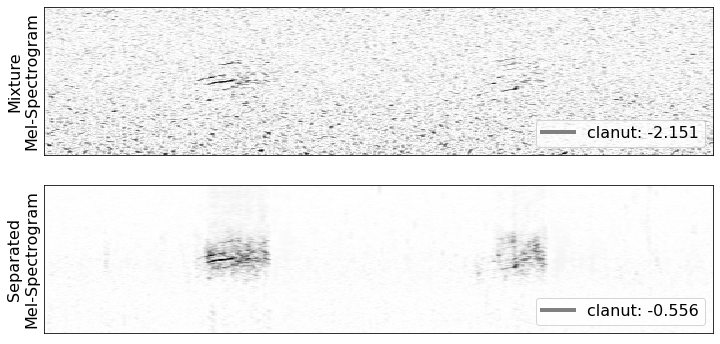}
  \end{subfigure}
  \hfill \vline \hfill
  \begin{subfigure}[b]{0.3\textwidth}
    \centering
    \includegraphics[width=\textwidth,trim={1.49cm 0 0 0},clip]{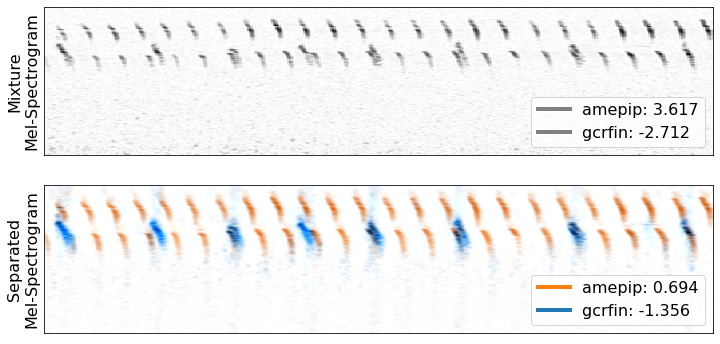}
  \end{subfigure}
  \hfill \vline \hfill
  \begin{subfigure}[b]{0.3\textwidth}
      \centering
      \includegraphics[width=\textwidth,trim={1.49cm 0 0 0},clip]{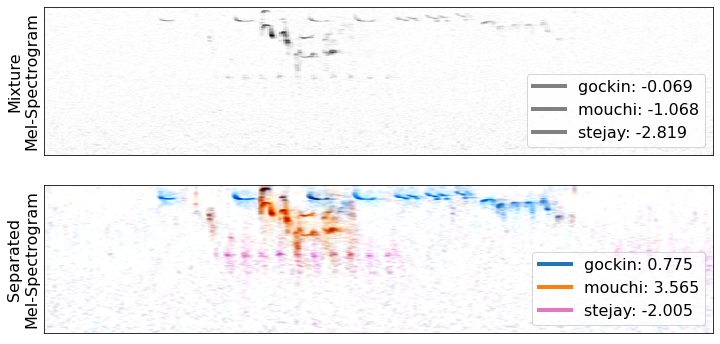}
  \end{subfigure}
  \caption{Separate+classify examples. Top plots show PCEN mel-spectrograms of original audio, and bottom plots show PCEN mel-spectrograms for separated audio, where separated channels are displayed with color-coding. The legend gives the ensemble logits for the ground-truth species. Left: a Clarke's Nuthatch in the High Sierras dataset, illustrating that simple noise suppression in the separated channel often improves logits for isolated calls with low SNR. Center: a challenging two-source example from the High Sierras dataset. Right: three-species separation from the Caples dataset.}
  \label{fig:examples}
  \vspace{-10pt}
\end{figure*}

\subsection{Evaluation Datasets}

We examine three separate evaluation datasets, using two distinct species sets covering upstate New York and the Sierra Nevada mountains in California. 

The \emph{Sapsucker Woods Dataset} (SSW) is the test set from the 2019 BirdClef competition \cite{birdclef2019} and contains 40k 5-second segments recorded in Ithaca, New York. The dataset contains 70 species, all appearing at least five times in the dataset. Audio was recorded by SWIFT Automated Recording Units (ARUs) and hand annotated by experts at the Cornell Lab of Ornithology.  We include the SSW dataset to allow some comparison with existing methods.

The \emph{High Sierras Dataset} (HSN) was collected at high elevation in the Sierra Nevadas and is described in the BirdClef 2021 competition \cite{birdclef2021}. After removing regions of silence from the annotated recordings, this dataset contains 4928 labeled 5-second segments, with 18 species appearing, of which 16 have at least 5 vocalizations. The annotated recordings were made between 4 and 8 AM, encompassing the dawn chorus.

The \emph{Caples Dataset} (CAP) is from the Caples Watershed near Lake Tahoe in California. Audio was collected from 80 points in 2017 and 2018 using ARUs as part of a project to study the effect of prescribed burning on biodiversity. To create the evaluation set, volunteers at the California Academy of Sciences chose 3-minute segments to annotate, providing time segmentation and confidence for labels. Low-confidence labels and regions of silence were discarded for evaluation purposes. The final dataset contains 2944 labeled 5-second segments, with 42 species. Of these, 36 species have at least 5 vocalizations.

\subsection{Evaluation Metrics}

We use a collection of metrics to understand model quality.

\emph{Class-averaged mean average precision} (CMAP) and \emph{label-weighted label-ranking average precision} (lwlrap) are closely related metrics useful for multi-class multi-label contexts, that generalize the mean reciprocal rank statistic for multi-label contexts. CMAP was the target metric for many BirdCLEF competitions \cite{birdclef2019}, and lwlrap was the target in the DCASE 2019 audio challenge \cite{dcase2019}. In brief, CMAP is the mean of the per-class precision scores, and lwlrap is the mean of the per-example precision scores. For species with few observations, per-class precision is very noisy, so we compute CMAP using only  species with at least five observations in a given dataset. Note that lwlrap reflects class imbalances in the dataset.

The \emph{$d'$ sensitivity index} is a re-scaling of AUC (treating every label as an independent binary classification problem with a common threshold). Define $d' = \sqrt{2} Z(\mathrm{AUC})$, where $Z$ is the CDF of the standard Gaussian distribution. It measures the overall quality of separation of positive and negative labels.

Finally, \emph{top-1 precision} is not a multi-label metric but is still helpful for evaluating model quality.

\subsection{Model Training Details}

The separation model is trained on a set of 704 species, including the New York and Sierra Nevada species and additional South American species included in the BirdCLEF 2019 challenge, with 6-second clips sampled at 22.05 kHz.  The learned filterbank uses a 1 ms window size, 0.5 ms hop, and 256 coefficients, and the model produces $M=4$ separated output channels.

Two separate ensembles of classifiers for the Sierra Nevadas (trained on a set of 89 species) and New York (trained on a set of 94 species). Input mel-spectra have a frame rate of 100 Hz, a frame-length of 0.08 seconds, a frequency range of 60 to 10,000 Hz, and 160 output channels.
Individual classifiers are trained for 200k training steps, using the Adam optimizer, with a learning rate of 0.01 and batch size 64. Each model trains in about 3.5 hours using 32 v2 TPUs. Longer training results in degraded evaluation CMAP. For the Xeno-Canto training examples, we set loss to zero for any species in the background labels to avoid punishing positive identifications.

\subsection{Separation Results}

To demonstrate the usefulness of training the separation model on a matched domain of audio data, we compare the separation model described in section \ref{ssec:sep_overview} trained on Xeno-Canto training data to a similar separation model trained with MixIT on 5800 hours of raw audio from AudioSet \cite{AudioSet}. The evaluation data consists of 3856 MoMs created from 6-second clips of held-out Xeno-Canto evaluation data. To measure performance we use the MoMi score \cite{wisdom2020mixit}, which is the mean \emph{scale-invariant SNR} \cite{LeRoux2018a} improvement (SI-SNRi) over estimated mixtures. For each example, the two estimated mixtures are computed by finding the optimal assignment of each separated source to one of the two reference mixtures and summing these sources together.

The closest pre-trained AudioSet model that was available for comparison has $M=4$ outputs, is trained on 16 kHz audio, and uses a learnable basis window length of 2.5 ms. To do a fair comparison, we also trained Xeno-Canto models with 2.5 ms window length instead of 1 ms, and we evaluated the 22.05 kHz model outputs by downsampling them to a sample rate (SR) of 16 kHz and scoring with 16 kHz references. Table \ref{tab:sep} shows the performance of Xeno-Canto-trained models compared to the pre-trained AudioSet model.
\begin{table}[h]
    \centering
    \caption{Separation results on Xeno-Canto MoM eval set.
    }
    \vspace{-5pt}
    \scalebox{0.9}{\begin{tabular}{|llll|r|}
    \hline
         Train data & Input SR & Eval SR & Window & MoMi  \\
         \hline
         AudioSet & 16 kHz & 16 kHz & 2.5 ms  & 4.4 dB \\
         Xeno-Canto & 22.05 kHz & 16 kHz & 2.5 ms  & 9.6 dB \\
         Xeno-Canto & 22.05 kHz & 16 kHz & 1.0 ms  & 10.4 dB \\
         Xeno-Canto & 22.05 kHz & 22.05 kHz & 1.0 ms  & {\bf 10.5 dB} \\
         \hline
    \end{tabular}}
    \label{tab:sep}
\end{table}
These results demonstrate that training on matched data for separation, rather than general audio data from AudioSet, does indeed provide a better separation model for this application, and provides a boost of over 5 dB in terms of MoMi on 16 kHz evaluation data. Decreasing the window size from 2.5 ms to 1.0 ms further improves performance by almost 1 dB.

\subsection{Comparison to Previous Classification Results}

 In the BirdCLEF 2019 \cite{birdclef2019} competition, the best model achieved a CMAP of 0.231 on the SSW dataset, which rose to 0.407 when trained with strongly-labeled data from the SSW validation set. Our classification models have not been trained on the validation set but have a somewhat larger training set, with up to 250 recordings per species, instead of the 100 per species in the BirdCLEF challenge. The New York classifiers achieve an average a CMAP of 0.242 individually and 0.304 as an ensemble of five.


\begin{table}[ht]
\centering
\caption{Comparing Separate+Classify methods.  ``Mix Only'' scores are the ensemble classifier scores for the raw audio. ``Separation'' scores are for the max probability for each species across the 4 separated audio channels. ``Mix+Separation'' takes the max species probability over the separated channels and the original audio. 
Classification is performed by an ensemble of EfficientNet-B0 models. \label{tab:sep_class_compare}}
\vspace{-5pt}
\scalebox{0.9}{
\begin{tabular}{|l|cccc|}
\hline
                          & \multicolumn{4}{|c|}{\textbf{Sapsucker Woods}} \\
                          & \scalebox{0.9}{CMAP}    & \scalebox{0.9}{lwlrap} & $d'$ & \scalebox{0.9}{Top-1} \\\hline
Mix Only                  & 0.304   & 0.431   & 1.117   & \textbf{0.398}\\
Separation                & 0.268   & 0.413   & 1.116   & 0.360\\
Mix+Separation            & \textbf{0.306}   & \textbf{0.441}   & \textbf{1.123}   & 0.397\\\hline
                          & \multicolumn{4}{|c|}{\textbf{Caples}} \\\hline
Mix Only                  & 0.334   & 0.569   & 1.144   & 0.496\\
Separation                & 0.327   & 0.581   & 1.154   & 0.506\\
Mix+Separation            & \textbf{0.341}   & \textbf{0.590}   & \textbf{1.155}   & \textbf{0.517}\\\hline
                          & \multicolumn{4}{|c|}{\textbf{High Sierras}} \\\hline
Mix Only                  & 0.527   & 0.531   & 1.149   & 0.432\\
Separation                & 0.548   & 0.548   & 1.149   & 0.448\\
Mix+Separation            & \textbf{0.560}   & \textbf{0.560}   & \textbf{1.153}   & \textbf{0.451}\\\hline
\end{tabular}
}
\label{table:sepclasseval}
\vspace{-10pt}
\end{table}


\subsection{Discussion of Results}
\label{subsec:discussion}

As shown in Table \ref{tab:sep_class_compare}, the evaluations show nearly uniform improvement of uncalibrated classification metrics across multiple datasets from combining the separation and classification models. Also, including the original noisy mixture as an additional channel gives better results than using the separated audio alone. Table \ref{tab:classifier_ablations} shows the results of an ablation study on the classifier, demonstrating the relative contribution of various components and architectures.

Three examples are shown in Figure \ref{fig:examples}, and more are available online\footnote{\url{https://bird-mixit.github.io}}. On examination of specific examples, we observed several benefits of separation.
    First, in many cases, isolated (non-overlapping) vocalizations with very low SNR obtain much better probabilities after removal of background noise (Figure \ref{fig:examples} left). The weakly labeled training data contains many unlabeled low SNR vocalizations that the model is therefore trained to ignore. After separation, PCEN processing brings these vocalizations clearly into the foreground.
    Second, mixtures of a common species and another species will often lead to a low probability for the common species, which improves after applying separation. This may be due to the common species appearing often in training recordings for other species, leading to lower probabilities.
    Third, our original motivation was to separate complex scenes with overlapping vocalizations, such as the dawn chorus. We find that the model is able to successfully separate overlapping vocalizations, even in many cases where frequency ranges of the vocalizing birds are overlapping (Figure \ref{fig:examples} center, right).

\begin{table}[t]
\centering
\caption{Classifier Ablations. For each classifier, we report the mean metrics for five individual models. We also provide the EfficientNet B0 Ensemble scores for comparison. 
         \label{tab:classifier_ablations}}
\vspace{-5pt}
\scalebox{0.9}{
\begin{tabular}{|l|ccc|ccc|}
\hline
& \multicolumn{3}{|c|}{\textbf{Caples}}&\multicolumn{3}{|c|}{\textbf{High Sierras}} \\
                        &\scalebox{0.9}{CMAP}& \scalebox{0.9}{lwlrap}& \scalebox{0.9}{Top-1}
                        & \scalebox{0.9}{CMAP}& \scalebox{0.9}{lwlrap}& \scalebox{0.9}{Top-1}\\\hline
B0 (Ensemble)           & 0.334   & 0.569     & 0.496          & 0.527   & 0.531  & 0.432\\\hline    
B0 (Mean)               & 0.284   & 0.517     & 0.458          & 0.479   & 0.483  & 0.392\\          
-Taxo Loss              & 0.272   & 0.491     & 0.434          & 0.465   & 0.469  & 0.376\\          
-Lowpass                & 0.283   & 0.467     & 0.423          & 0.401   & 0.398  & 0.319\\          
-Channel Norm           & 0.270   & 0.484     & 0.437          & 0.438   & 0.447  & 0.362\\\hline    
B0 (Mean)               & 0.284   & 0.517     & 0.458          & 0.479   & 0.483  & 0.392\\          
B1                      & 0.275   & 0.503     & 0.448          & 0.477   & 0.479  & 0.396\\          
B2                      & 0.268   & 0.490     & 0.437          & 0.486   & 0.477  & 0.390\\          
\hline
\end{tabular}
}
\label{table:classablate}
\vspace{-10pt}
\end{table}

The use of separation also produced some shortcomings.
    First, we often observe that the probability of the most prominent species in a given recording \emph{decreases} after separation. This may be due to loss of additional audio context. Including the original mixture with the separated channels mitigates this effect.
    Second, evaluation scores degrade if we apply classification only on the separated channels. This may be due to over-separation, domain shift, or loss of audio context.
    Third, over-separation occasionally occurs and sometimes isolates a few notes from a longer song. Isolated notes may resemble calls of another species, resulting in a misclassification. 


\section{Conclusion}
\label{sec:conclusion}

We have demonstrated the utility of training a domain-specific MixIT separator for bioacoustics through combination with a downstream classifier. There are clear extensions of this work to other bioacoustics problems, such as frog chorusing and marine soundscape analysis. We hope to find improved methods for combining the separation and classification systems. For example, applying separation to classifier training data may improve classifier quality. Targeted separation of sources into specific channels would also allow the creation of better \emph{acoustic indices} by separating bird, insect, and amphibian vocalizations into distinct channels.

\section{Acknowledgements}

We would like to thank Mary Clapp, Jack Dumbacher, Durrell Kapan and the California Academy of Sciences for the Caples and High Sierras datasets and for extensive qualitative feedback on classifier quality. We are also grateful to Stefan Kahl and Holger Klinck and the Center for Conservation Bioacoustics at the Cornell Lab of Ornithology for the Sapsucker Woods dataset. All models were trained with data from Xeno-Canto and the Macaulay Library; this work would not be possible without them.

\vfill\pagebreak


\bibliographystyle{IEEEbib}
\balance
\bibliography{refs}

\end{document}